# Topological Quantum Materials for Energy Conversion and Storage

Huixia Luo[1]*, Peifeng Yu[1], Guowei Li[2,3]*, Kai Yan[4]*

[1]School of Materials Science and Engineering, State Key Laboratory of Optoelectronic Materials and Technologies, Key Lab of Polymer Composite & Functional Materials, Guangzhou Key Laboratory of Flexible Electronic Materials and Wearable Devices, Sun Yat-Sen University, 135 Xingang Xi Road, Guangzhou 510275, China

[2]CAS Key Laboratory of Magnetic Materials and Devices, and Zhejiang Province Key Laboratory of Magnetic Materials and Application Technology, Ningbo Institute of Materials Technology and Engineering, Chinese Academy of Sciences, Ningbo 315201, China

[3]University of Chinese Academy of Sciences, 19 A Yuquan Rd, Shijingshan District, Beijing 100049, China

[4]School of Environmental Science and Engineering, Sun Yat-Sen University, 135 Xingang Xi Road, Guangzhou 510275, China

E-mail: luohx7@mail.sysu.edu.cn; liguowei@nimte.ac.cn; yank9@mail.sysu.edu.cn.

Abstract | Topological quantum materials (TQMs) have symmetry protected band structures with useful electronic properties that have applications in information, sensing, energy, and other technologies. In the past 10 years, applications of TQMs in the field of energy conversion and storage mainly including water splitting, ethanol electro-oxidation, battery, supercapacitor, and relative energy-efficient devices have attracted increasing attention. The novel quantum states in TQMs provide a stable electron bath with high electronic conductivity and carrier mobility, long lifetime, and determined spin states, making TQMs an ideal platform for understanding the surface reactions and looking for highly efficient materials for energy conversion and storage. In this Perspective, we present an overview of the recent progress regarding topological quantum catalysis. We describe the open problems, and the potential applications of TQMs in water splitting, batteries, supercapacitors, and other prospects in energy conversion and storage.

# Introduction

Topological quantum materials (TQMs) host symmetrically protected, high mobility electronic states[1-6]. These features make them attractive for a range of applications – most commonly discussed are their potential for spin-related information storage and processing, but TQMs can also be useful for energy conversion and storage. In the past decade, there has been a growing interest in investigating the potential use of TQMs in energy-efficient applications[7-21] such as water splitting, thermoelectricity, battery, supercapacitors, and ethanol electro-oxidation. However, the interaction between topological properties and the reaction processes still needs deeper understanding. Further study of topological properties is needed from both a chemistry and physics perspective to uncover the relationship between topological band structures and energy conversion or storage, for the design of high-efficiency heterogeneous catalysts for energy conversion such as water splitting and fuel cells.

In this Perspective, we present a brief overview of the recent development and advances in the use of various TQMs for energy conversion and storage technology. We propose that TQMs could be a new category of energy material for the various utilizations in photo-electrocatalytic water splitting, $CO_2$ reduction, CO oxidation, battery, supercapacitors, and beyond. We begin by giving a general diagram for catalysis and energy conversion, followed by a review of the recent theoretical research and experimental progress of TQMs for energy conversion and storage. We focus on the applications of TQMs in water splitting, battery and survey the current experimental and theoretical progress. The TQMs used for selective hydrogenation, oxidation of biomass-derived alcohols, storage of hydrogen, and beyond also has been briefly surveyed. In the end, we will give the outlook for the challenge of current topological catalysis and the potential new applications of TQMs. In the Supplementary Information, we provide a brief discussion of the application of TQMs in the thermoelectricity field. For a complimentary review on topological thermoelectric we direct readers to REF.[7] and for a review of the topological insulators and thermoelectric, see REF.[22] and REF.[23].

# Water splitting

Hydrogen is a promising energy carrier due to its large energy density without carbon emission, which is believed as a solution to environmental problems and energy crisis[24-27]. Thus, water electrolysis to produce high-purity hydrogen as the alternative clean fuel to traditional fossil fuels has attracted increased research attention[28,29]. This process consists of hydrogen evolution reaction (HER) and oxygen evolution reaction (OER), two half-reactions, which can be driven by electricity or light with suitable energy[24]. HER and OER involve two-electron and four-electron-transfer process which often requires high overpotential ($\eta$) to overcome the large energy barrier[25,26]. Considerable research efforts have been devoted to developing efficient electrocatalysts to improve the electrochemical conversion efficiency and decrease the overpotential of HER as well as OER process. Traditional strategies including defect engineering, nano-technology, and single atomic technology are proved to be effective in the search for low-cost and high-efficient catalysts [30-33]. Most key steps in water splitting, such as mass adsorption, electron transfer, and desorption take place on the catalyst surface. Unfortunately, it is still a tough job to tell the origin of catalytic activities and then replace the state-of-the-art noble metal catalysts.

For a typical redox reaction at the crystal surface, electrons are transferred either from the catalysts to the reactant molecules (reduction reaction), or in the reverse direction (oxidation reaction). Thus, the electronic conductivity of catalysts is important because electrons should be passed through the bulk phase and then to the external circuit. In addition, high mobility is necessary to avoid the trapping and recombination of carriers (**BOX 1**). However, the traditional catalysts suffer from surface contamination during the catalytic reaction or poor electronic conductivity, leading to low catalytic activity or high reaction barriers[27]. To overcome these problems, the search for novel high-effective catalysts is an urgent need for water splitting.

The topologically protected surface states, giant electron motilities, and chiral surface states of TQMs, make them ideal candidates for catalysis reactions such as water splitting[27]. The TQMs with robust TSSs can resist perturbations such as defects, a small amount of doping, and surface modifications. Moreover, the TQMs with unique topological electronic band structures can stably supply the topological electrons into the studied catalysts as a robust electron bath. Hence, the TQMs may help to overcome the aforementioned issues for the traditional catalysts. In addition, the out-pointing of $p_z$ components of the $p$ orbital and the $d_{z^2}$ components of the $d$ orbital tend to form covalent bonds with the surface molecules through σ or π bond[34,35]. Once the electron transfer channel is established, reduction or oxidation reactions can be initiated by applying an electric potential or light illumination. The presence of the surface states has great importance in mediating the interfacial charge transfer and resulting in different reaction pathways depending on the applied potential.

## Topological materials for HER

### *Topological insulators*

The research on the catalytic behaviour of TQMs started from the family of 3D topological insulators (TIs) based on bismuth chalcogenides ($Bi_2Se_3$ and $Bi_2Te_3$)[2-4,36]. The response of topological surface states (TSSs) at the surface of $Bi_2Se_3$ the deposition of metals like Au, Ag, Cu, Pt, and Pd was investigated using DFT caluclations[37] and improved adsorption energies of oxygen were seen, which indicated the importance of TSSs for the tailored adsorption behaviour. However, this result alone does not suggest optimized catalytic efficiencies because only the metals with original weak reactivity can attain a positive effect from TSSs. This was firstly experimentally and theoretically confirmed by the photocatalytic hydrogen studies on nanostructured $Bi_2Se_3$ and $Bi_2Te_3$ catalysts[8]. Benefitting from the high mobility, it was found that the photoexcited electrons from the TSSs stay for a longer time than their counterpart compounds (such as $Bi_2Se_3$, $Bi_2Te_3$, $Bi_2Te_2Se$), which leads to efficient electron-hole separation, then giving rise to the high-effective photochemical HER performance[8]. The effect of TSSs can be maximized by making highly-crystalline thin films with desired exposing crystal surfaces. For example, $Bi_2Te_3$ and SnTe thin films exhibited remarkable intrinsic HER activities and maintained high stability against slight surface oxidation and defects[38,39]. More interesting, the TSSs could not only interact with the adsorbed hydrogen directly, but they can also change the catalytic performance of the deposited catalysts[40]. The Pd (20 nm)/SnTe (70 nm) heterostructure fabricated by molecular beam epitaxy displays much better HER performance and turnover frequency of per-Pd-site in comparison with the Pd (20 nm) thin film and is even superior to the commercial Pt foil. Such a good HER activity is the result of the transfer of electrons from both the Pd surface and the adsorbed H atoms to the TSSs adhered to the SnTe (001) underlayer, which leads to the weaker Pd-H binding strength and more favourable hydrogen adsorption free energies[41]. These findings also indicate that it is not technically a requirement for the direct bonding between the TSSs and adsorbates.

*Topological semimetals*

A hindrance to the application of TIs as catalysts is the low chemical stability of metal chalcogenides. Although the topological surface properties are protected by their symmetry, they can be destroyed under harsh catalytic conditions. The upmost few layers of air-exposed $Bi_2Se_3$ TIs at ambient temperature could be reconstructed and can be even completely vanished in the acidic/alkaline electrolytes. This will significantly decrease the contribution of TSSs in the catalysis reactions[42,43]. In addition, the bulk phase of TIs is characterized by a bandgap like the conventional semiconductors, albeit the bandgaps are generally not big, and the crystal surfaces are metallic[44]. The low conductivity of the bulk phase will increase the charge transfer resistance or lower the charge-hole separation efficiencies when used as catalysis, and thus are detrimental to the catalytic kinetics. With this in mind, topological semimetals received increasing research interest because of their high chemical stability and relatively high electronic conductivity. Topological semimetals (TSMs), mainly including Weyl semimetals (WSMs), Dirac semimetals (DSMs) and nodal line semimetals (NLSMs) appear at the borderline between topological and trivial insulators and have become the major members of TQMs[45,46]. It has been noticed that the existence of conducting TSSs cannot assure good catalytic performances. The weak interaction between TSSs and adsorbates will lead to low coverages and slow catalytic reaction kinetics[38,47,48]. Potential TQMs catalysts need to be able to boycott turbulence under surface modification while having high electronic conductivity and carrier mobilities for the fast electron transport during the catalytic reaction process. The robust TSSs of TSMs with high electronic conductivity and carrier mobilities may promote the catalytic activity.

Application of the TSMs with high carrier mobilities for water splitting was firstly realized in the photocatalyst materials, transition-metal monopnicitides WSMs (TaP, NbP, TaAs and NbAs)[49]. The carrier mobility of these transition-metal monopnicitides is in the order of $10^5$ cm$^2$ V$^{-1}$ s$^{-1}$ (Table 1). Such high mobility boosts fast transfer between TSSs and surface adsorbates, leading to high catalytic activity[50]. Subsequently, Pt-containing TSMs (PtSn$_4$, PtAl, and PtGa) with high carrier mobilities were also documented to show superior HER catalytic performance, even exceeding those of some noble metal-containing catalysts[13,16]. In addition, group VIII noble-metal dichalcogenides (PtSe$_2$, PdTe$_2$, and PtTe$_2$) have recently been theoretically predicted and experimentally confirmed to be DSMs and host type-II Dirac fermions[51-54].

It also has been theoretically proposed that the double vacancies in Se or Te (DV$_{Se}$ or DV$_{Te}$) of DSMs and boron-substitution can adjust the interplay between H and basal plane and then boost the HER performance of these monolayer noble-metal dichalcogenides[55]. In particular, the Gibbs free energy for hydrogen adsorption is decreased from 0.94 eV for the pristine PdSe$_2$, to only 0.02 eV with 2 % boron substitution. This corresponds to fourteen orders of magnitude increase in the exchange current density after doping. In addition, it is well-known that the topological band structures are robust against surface modification such as elemental doping, defects, surface atomic reconstruction, surface oxidation, and molecules adsorption. For example, adding Na vacancies to the DSM Na$_3$Bi Shifted the Fermi level upward, but the bulk Dirac cone persisted despite heavy surface modification[56]. Similarly, the TSSs of the TI Bi$_2$Se$_3$ is stable against N$_2$ and air exposure, and even oxidation[57]. Therefore, the interaction between the variations of TSSs and the enhanced HER catalytic activities of these noble-metal dichalcogenides is worth further study.

It should be noted that the traditional Pd-based electrode is not an ideal electrocatalyst for HER due to the formation of metal hydride[58,59]. The presence of the robust TSSs seems to not only reduce the metal hydride effect but also enhance the HER catalytic activities and stabilities, as seen in the case of the topological and superconducting α-BiPd catalyst, which shows a striking improvement of the catalytic activity and stability. It is found that α-BiPd shows a shorter Pd-Pd bond in comparison with the pure Pd metal, which leads to the large density of states with conduction electron at the Fermi level[60,61].

The cost and abundance of the catalysts in practical usages should be also taken into consideration besides the catalytic activity, selectivity, and stability. Recently, the noble metals-free TSMs have also been found to show excellent HER catalyst candidates[62-64]. Especially, the low-dimensional transition metal dichalcogenide (TMD) system is one of the most extensively studied topological semimetals due to its low dimensional structure and rich topological properties. For example, the bulk orthorhombic T$_d$-MoTe$_2$ TMD is type-II WSM[65-69], while the monolayer and few-layers two-dimensional (2D) 1T'-MX$_2$ (M = W, Mo; X = Te, Se, or S) have nontrivial Z$_2$ topological invariants with quantum spin Hall (QSH) effect feature[70,71], although bulk 1T'-MoS$_2$ is semiconducting with a fundamental gap (E$_g$ ≈ 0.08 eV). Early photocatalytic measurements of the topological 2D 1T'-MoTe$_2$ TMD provide support for the topological heterogeneous catalysis in water splitting[62]. Later work finds that topological 2D 1T'-MoTe$_2$ nanosheets were also good for electrocatalytic hydrogen generation, where both the single-crystal and polycrystalline 2D 1T'-MoTe$_2$ nanosheets display remarkable performance for hydrogen evolution[64].

Size reduction can enhance the catalytic properties of TQMs, for example, an improvement in the HER activity has been observed in the polycrystalline powder materials with smaller sizes compared to the single crystals with larger sizes[62,64]. Hence, it can be expected that the catalytic performance of WSMs can further be promoted largely by decreasing the particle size, for instance, by constructing them as nanoparticles until the bulk remains metallic. Recently, noble metals-free $VAl_3$-based topological semimetals with large carrier mobility at room temperature were also verified to be highly-effective and acid-stable topological catalysts, where the direct transfer of electrons from the occupied robust TSSs to the adsorbed H boost the electrochemical HER efficiency[63]. These results affirm that the enhancement of the HER activity in these noble metals-free TQMs is a result of the TSSs and the high mobility.

*d-band topological materials*

Although great progress has been made in previous works on employing TQMs in HER, most of these TSSs are derived from the *sp* band of topologically trivial metal catalysts rather than the *d* band. However, based on the traditional adsorption mechanism, the *d* band figures prominently in determining catalytic efficiency[72]. This is because only the *d* orbitals could form strong bonding with the surface adsorbates. Otherwise, the adsorption is not energetically favored and leads to low reactant coverage, which will decrease the intrinsic activities of the catalysts.

Therefore, TQMs in which the TSSs are derived from the *d* band may be beneficial to the catalytic activity. The observation of unconventional topological fermions (such as chiral fermions, topological non-Fermi liquid, Dirac nodal line, large topological Fermi arcs, and so on) in the group of transition-metal monosilicides has received increasing research attention[73-78]. It is interesting to see that the TSSs are mainly derived from the transition metal *d* orbitals. More importantly, they are also accompanied by large topological non-trivial energy windows and a large density of states around the Fermi level[79,80]. This suggests an enlarged contribution of TSSs in the catalysis reactions. One example of a transition-metal monosilicide is CoSi., which belongs to the space group of $P2_13$ with a cubic structure, where each Co atom is connected with six Si atoms and vice versa (**Fig. 1a**). Based on the first principle calculations prediction, and supported with later ARPES experimental evidence, it is found that peculiar massless fermionic excitations with nonzero Berry flux such as spin-1 excitations with threefold degeneracy and spin -3/2 Rarita-Schwinger-Weyl fermions coexist in the transition metal silicide CoSi (**Fig. 1b**) (REFS[75,77]).

Recently, a group of transition-metal monosilicides ($M$Si, $M$ = Co, Ni, Ti, Fe, Mn, Rh, and Ru) has been synthesized by the arc-melting method within several minutes and the easily fabricated $M$Si polycrystalline electrodes display remarkable activities in the HER process with overpotentials of 34 ~ 54 mV at the current density of 10 mA cm$^{-2}$ as well as low Tafel slopes of 23.6 ~ 32.3 mV dec$^{-1}$ (see REF.[11]). These are similar or even superior to 37 mV, 26.1 mV dec$^{-1}$ of 20 wt% Pt/C (**Fig. 1c**). The theory calculations confirm the outstanding activities are closely related to hydrogen coverages (~ 100 %) on several low surface energy facets. When compare the free energy difference |ΔG| of the previously reported topological catalyst with the commercial Pt in a volcano plot, the free energy difference |ΔG| of transition-metal monosilicides is close to the commercial Pt catalyst sitting near the peak of the volcano plot (**Fig. 1d**). The subtle deviation of the hydrogen adsorption energy and the free energy difference |ΔG| of these $M$Si are mainly contributing to the transition metals filled by the different $d$ orbital.

Furthermore, catalytic properties can be tuned and enhanced by various alloying combinations, which opens a new avenue to use low-cost silicon-based candidates for HER. This is maybe a good sample for the explanation of the enhancement of the HER activity by the combination of the nontrivial TSSs and the conditional $d$ band theory (see REF.[11]). However, the link between robust TSSs and the high HER catalytic activity still needs to be fully understood. Recently, it is found that topological charge carriers in NiSi WSMs with hybrid Weyl states and long surface Fermi arcs close to the Fermi level indeed take part in the HER process (**Fig. 1e**). In the comparison of the surface band structure of NiSi without and with H adsorption, four Weyl node points shift down on the NiSi surface with H atoms and the Fermi arcs shifts toward lower energy after H adsorption, indicating that the charge transfer takes place on its surface and the charge exchange between NiSi surface and H atom in the HER process (**Fig. 1f**)[81]. These results further indicate that it is highly possible to design HER catalysts without noble metals via topological engineering of the band structure.

Topological materials for OER

Compared to the topological HER catalysts, topological OER catalysts have been less studied so far, although numerous excellent OER catalysts have been discovered by introducing applying strain, vacancies, defects, doping, or tuning transition-metal coordination. However, as these strategies rely on extrinsic disturbances, they are closely paired with structural instability, leading to the instability of the OER activities. In the OER process, the adsorption of diverse reaction intermediates and electronic structures has an important effect on regulating the adsorption processes. Based on the Shao-Horn (SH) principle, the orbital filling, $e_g$ affects the binding energy of oxygen intermediates to the catalyst surface and changes the OER performance; hence we can tune the $e_g$ of metal sites to optimize OER activities[82] (**Fig. 2a**). Recently, ternary Heusler compounds have been successfully documented to be excellent OER catalysts by precisely controlling $e_g$ (**Fig. 2b**) (REF.[83]). Another prominent example is the TSM $Co_3Sn_2S_2$ bulk single crystals with high electrical conductivity and robust surface states at room temperature. The TSSs are mainly derived from the Co 3*d* orbitals in the Kagome lattices and are just located above the Fermi level (**Fig. 2c**). The bulk single crystals with TSSs display eminent OER catalytic activity and even surpass the Co-based nanostructured catalysts, although the specific surface area of the bulk single crystal is much smaller. The high efficiency of the bulk single crystals can be explained by the half-filled Co $e_g$ orbital in the TSSs, which are pointing towards the *p* orbital of the adsorbed hydroxide ions. Such an $e_g$ - *p* orbital overlap will strengthen the bonding between the active sites and the adsorbates and make the electron transfer more efficiently through the bonds, thereby resulting in the high OER performance of the Co$_3$Sn$_2$S$_2$ bulk single crystals and comparable to the state-of-art RuO$_2$ catalysts (**Fig. 2d**) (REF.[12]). These findings highlighted the significance of using TSSs as the catalytic active sites in the design of high-efficient catalysts.

# Batteries and supercapacitors

### Lithium-ion batteries

TQMs with symmetry-enforced Dirac states provide robust carriers for high electronic conductivity and can improve battery efficiency. Many topological carbon-rich compounds are theoretically predicted to be good anodes for lithium-ion batteries (LIBs) when compared with conventional carbon compounds such as graphite, carbon nanotubes or porous carbon nanofibers (See Table 2). In fact, these carbon compounds have been extensively used as anodes for LIBs owing to their excellent reversibility, low environmental impact, and cost[84]. In particular, porous carbon compounds are useful because of their big specific surface area, which can offer abundant binding sites for Li-ions and thus favours high-capacity LIB anodes. The traditional porous carbon candidates used in LIBs are commonly amorphous[85], where the disordered holes and defects have an impact on the diffusion of Li-ions and the irreversible capacity is usually high[86].

There are numerous theoretical discussions on the use of topological semi-metallic porous carbon materials as potential LIBs anodes, but the influence of TSSs needs to be further studied experimentally[87-91]. It has been found that these all-carbon-based porous 3D topological carbon compounds with inherently ordered porosity possess higher electrical conductivity[92,93] in comparison with the conventional disordered porous carbon compounds and hence are beneficial for anode components. The use of the topological semi-metallic porous carbon materials for LIBs has its roots in the pioneering work in REF. 87. From this work, the all-carbon-based porous body-centered orthorhombic C$_{16}$ (bco-C$_{16}$) as a high-effective LIB anode candidate could be understood as a consequence of a novel topological NLSM bco-$C_{16}$ (**Fig. 3a, b**) with intrinsic high electronic conductivity[89] and 1D migration channels (**Fig. 3c**). The ion diffusion channels in bco-$C_{16}$ are immune to the compressive and tensile strains during charging/discharging, which leads to a larger specific capacity (the maximum Li concentration, Li-$C_4$) than that of the usually employed graphite anode (Li-$C_6$). Furthermore, the energy barrier declines with enhancing Li insertion and can achieve 0.019 eV at large Li-ion content; the average voltage is very low (0.23 V), and the volume change during the test keeps pace with that of graphite[87] (**Fig. 3d**). bco-C16 has a larger volume expansion (13.4 %) than that of the usually utilized graphite anode (10 %).

Later work[94] proposes a porous 3D nodal line semimetal carbon (termed HZGM-42) as a potential anode compound for LIBs, which shows a large theoretical specific capacity (637.71 mAh $g^{-1}$) and a small energy barrier (0.02 eV) for Li-ion transport along the ordered 1D channels as well as a subtle volume change (2.4%) during charging and discharging processes. This performance is the result of the larger pore size in topological porous carbon HZGM-42 with hexagonal symmetry compared to the interlayer distance in graphite and intrinsic superior electrical conductivity in comparison with the disordered carbon, metal oxide, metal sulfide, and selenide anodes[95-97].

One year after the discovery of porous nodal line semimetallic carbon, monoclinic $C_{16}$ (m-$C_{16}$)[98], it is noticed that m-$C_{16}$ exhibited superior performance as an anode compound for LIBs to that of the previously reported bco-$C_{16}$. TQM m-$C_{16}$ possesses a similar specific capacity (558 mAh $g^{-1}$) but with a smaller volume change (3.6%) during the charging/discharging performance and a lower Li-ion transport energy barrier (0.25 eV) when compared to corresponding values of 558 mAh $g^{-1}$, 0.53 eV and 13.4% in bco-$C_{16}$ (REF.[88]). The Li atoms contribute nearly all valence electrons to TQM m-$C_{16}$, resulting in a strong ionic bonding with the substrate and consequent ascendant performance.

With the prediction of the topological porous carbon as LIB anode materials, it is natural to ask whether they can be realized experimentally. There has been a proposal to use a high-pressure synthesis technique with the universal structure database to seek a topological semimetal porous carbon as the expected anode. It is found that an orthorhombic phase $LiC_6$ with uniform pores can be acquired at 30 GPa. In addition, the obtained carbon structure after removing the Li atoms is still a topological semimetal with an interlocked graphene network (IGN) and an inherent large electronic conductivity. A low Li-ion diffusion energy barrier and three orders of magnitude higher diffusion coefficient ($10^{-4}$ $cm^2$ $s^{-1}$) than that of graphite anode are observed at both small and large Li concentrations in the IGN. Besides, the volume changes during the Li insertion/extraction processes are as low as 3.2% and keep the same theoretical specific capacity as that of graphite anode[99].

## Sodium-ion batteries

An alternative to LIBs is sodium-ion batteries (NIBs), as sodium is abundant and inexpensive compared to lithium. Similar to the aforementioned LIBs cases, the 3D porous topological carbon material, tetragonal $C_{24}$ ($tC_{24}$) with two topological nodal surfaces crossing the whole Brillouin zone, recently have been predicted to be a potential anode compound for NIBs. $tC_{24}$ has a long cycling life, a high predicted capacity of 232.65 mAh $g^{-1}$, an appropriate average voltage of 0.54 eV, a low diffusion energy barrier for Na ions and a negligible volume change (~ 0.94%) during the charge/discharge processes[90]. There are three distinct voltage regions for varying Na content in $Na_xtC_{24}$ and the voltage maintains positive during the whole operation, indicating that the half-cell reaction can occur automatically to achieve the ultimate component ($Na_5C_{48}$) (**Fig 3e**). Therefore, the $tC_{24}$ anode can have a wholly reversible capacity of 232.65 mAh $g^{-1}$ (REF.[90]). The aforementioned 3D NLS porous carbon HZGM-42 is a potential anode candidate not only for LIBs but also for NIBs with high performance[100]. All these facts indicate that the all-carbon HZGM-42 with large energy density, excellent rate capability as well as good cycling stability can be used as a versatile anode candidate for both LIBs and NIBs.

In contrast to porous topological carbon materials, the topological silicon materials employed as anode materials for NIBs have attracted less attention, despite silicon and carbon being located in the same group in the periodic table. Recently, it has been found that the all-silicon NLSM consisting of interpenetrating silicene networks, is a potential material for NIBs anode due to its comparatively big storage voltage (1.35 V), inherent metallic property together with particularly small barrier energy (5.2 meV), and subtle volume change[101]. More impressively, hexagonal supertetrahedral-boron consisting of B4 cluster is first proposed to be a 3D porous topological metal with several kinds of spin-orbit-free burgeoning fermions (REF.[102]) and subsequently is found to be an anode compound for both LIBs and NIBs with an ultrahigh capacity of 930 mAh g$^{-1}$, low migration barrier of 0.15 and 0.22 eV and low volume changes of 0.6 % and 9.8 % for Li and Na ions, respectively[103]. These enchanting characteristics imply that 3D porous TQMs deserve further investigation for battery devices[104].

Recently, $Bi_2Te_3$ is utilized for sodium storage (**Fig. 3f**). Based on the *in-situ* XRD characterization, there exists a four-step crystallographic phase-change mechanism of $Bi_2Te_3$ during sodiation/dissociation processes. After that, the WSM $T_d$-$Wte_2$ was applied as the electrode material for NIBs, which delivers a 245 mAh g$^{-1}$ at 100 mA g$^{-1}$ (REF.[105]). The Na-based TQMs also provide new promising material candidates for NIBs. For instance, NaAlSi and NaAlGe which are dual double NLSs (those that have both type-I nodal lines and type-III nodal points) are theoretically predicted to be potential candidates for NIBs, due to their large proportion of Na ion, with a predicted reversible capacity of 476 and 238 mAh g$^{-1}$, respectively (see **Fig. 3e** in REF.[91]).

Furthermore, 3D DSMs, such as $Na_3Bi$, host 3D Dirac fermions that are immune to in situ surface K-doping[55]. Thus 3D DSMs have a theoretical capacity of 385 mAh g$^{-1}$ (REFS[106,107]). Besides, a topological phase transition from DSMs to a trivial band insulating compound by alloying has been observed in $Na_3Bi_xSb_{1-x}$ alloys, which is derived from the strength of the spin-orbit-coupling (SOC) by varying with Sb content. The topological state-driven from SOC in $Na_3Bi$ is very robust compared to anode material $Na_3Sb$. The presence of Dirac bands with high carrier velocities and the associated conductivity offer a favorable factor for battery performance[108].

There is a large number of existing materials are theoretically predicted or experimentally confirmed to be TQMs[4,6,36,109,110,111]. But, what kinds of TQMs are promising to be candidate battery materials? This question has been approached using machine learning theory to go through 7300 TQMs for promising NIBs cathodes, and then using DFT calculations to confirm them[110]. If a TQM cathode ensures large electrical conductivity, then large ionic conductivity is required for high-efficient activity. Building on this idea, in Ref 110, a descriptor-based machine learning model involving atomic radius, Pauling electronegativity, many valence electrons, first ionic energy, and radial basis functions and a supervised Gaussian Process Regression (GPR) model, is used to predict the Na-ion migration energy barrier for the topological NIB candidates. The data mining displays that the atomic radius plays an important role in the migration energy barrier, and the cations affect more significantly than anions for Na ion migration. This calculation provides a powerful and simple rule for further search for new topological battery cathode materials[110].

## Potassium-ion batteries

Potassium-ion batteries (KIBs) have also been considered as one of the possible substitutions for LIB due to the negative potential of the $K^+/K$ redox couple, the considerably adequate resources of K, and the practicable usage of low-cost aluminum (Al) current collector in KIBs. The discovery of inherent porosity-ordered channels and large electronic conductivity in the topological carbon compounds led to the proposal of a 3D porous NLSs carbon allotrope (named BDL-14) as a promising KIB anode. The topological BDL-14 shows superior properties of anodes to those of the anodes currently being considered[112].

## Aqueous zinc-ion batteries

Aqueous ZIBs (AZIBs) are also considered to be a promising alternative to the widely used LIBs in large-scale electrical energy storage systems because of the cheap, high safety, and atoxic nature of zinc. One significant factor that demands further development is finding electrode materials that store $Zn^{2+}$ ions with large invertibility and rapid kinetics[113]. WSMs with excellent electrical conductivity and high carrier density around the Fermi level are appealing candidates for AZIBs, benefiting from TSSs. The Weyl semimetal, $Co_3Sn_2S_2$, is recently proposed to be an AZIB cathode with a discharge plateau near 1.5 V. Furthermore, more $Zn^{2+}$ transfer channels and enhancing charge-storage kinetics as well as higher $Zn^{2+}$ storage ability are obtained by the introduction of Sn vacancies. The synergy benefits of leveraging TSSs and introducing vacancies have been realized in the $Co_3Sn_2S_2$ battery performance[114].

## Supercapacitors

Apart from ion batteries, TQMs also offer new opportunities in other high-efficient energy storage devices such as spin battery[17] and supercapacitors[115]. In the last 5 years, supercapacitors have gained incalculable attention owing to their large power density, rapid charge/discharge rates, and excellent cycling stability, all features show great potential for bridging supercapacitors and batteries. According to the different storage mechanisms, supercapacitors can be decided into two classifications: electrochemical double-layer capacitors involving reversible ion adsorption/desorption on the electrode-electrolyte interface and pseudo-capacitors, which depend upon the Faradaic redox reactions. Currently, layered 2D $Bi_2Se_3$ nanosheets are used for supercapacitors and exhibit a specific capacitance of 439 F $g^{-1}$ at 4 A $g^{-1}$ (REF.[116]). When $MoSe_2/Bi_2Se_3$ hybrid nanosheets are used for supercapacitors, they give capacitance of 1451.8 F $g^{-1}$ at 1 A $g^{-1}$ and maintain 750 F $g^{-1}$ at 20 A $g^{-1}$ (REF.[117]). Another example is the TMD tungsten ditelluride ($1T_d$-$WTe_2$), which has been first theoretically predicted[118] and subsequently experimentally documented to be a type-II Weyl semimetal[119-121], showing a promising advanced electrode used in all-solid-state pliable supercapacitors. In this case, $1T_d$-$WTe_2$ single-crystal bulk is exfoliated into nanosheets by liquid-phase exfoliation and then fabricated into air-stable films by drop-casting $1T_d$ $WTe_2$ nanosheets on arious substrates including Au/Ti coated PET substrate and Si/SiO$_2$ wafer, which acts as the electrode directly. The films are then assembled into all-solid-state pliable supercapacitors by using a gel electrolyte of poly(vinylalcohol)/H$_3$PO$_4$ (PVA/H$_3$PO$_4$) sandwiched between two films, which show superior performance to the commercial 4V/500 μAh thin-film LIB and 3V/300 μAh Al capacitor, accompanied by splendid flexibility and excellent cyclability[115]. These facts further verify that the TQMs offer new material platforms to explore the battery devices.

# Multi-electron transfer reactions

## Catalysis for $CO_2$ conversion

TQMs with TSSs are not only good candidates for single electron transfer reactions such as HER but also proved to be efficient in multi-electron processes including carbon monoxide (CO) oxidation[122], carbon dioxide ($CO_2$) electro-reduction[123,124], and selective hydrogenation of alkynes[125,126]. An early study employs gold-covered 3D $Bi_2Se_3$ as a prototype example for CO oxidation[122]. Based on first-principles DFT calculations, it is found that the TSSs severs as a highly-effective electron bath. The dissociation and adsorption of $O_2$ and CO molecules on a gold-covered $Bi_2Se_3$ structure can be promoted by raising the adsorption energies of these molecules. Later, the experimental study of the influence of TSSs on various surface reactivity in Pd-covered $Bi_2Te_3$ is reported which agreed well with the aforementioned theoretical predictions[127]. Moreover, the theoretical results imply that the TSSs from $Bi_2Te_3$ can tunnel through a thin Pd layer and remarkably boost the surface activity of Pd[127]. Bismuth, which is a topological crystalline insulator, can be dissolved under HER conditions and is not suitable for water electrocatalysis. But the Bi nanostructures with exposed surface states are one of the most efficient catalysts for $CO_2$ reduction which exhibited a small overpotential and high selectivity of Faradic efficiency over 92.2% in a wide partial current density range from 9.8 to 290.1 mA cm$^{-2}$ (REF.[128]). The topological surface states preferentially adsorbed *OCHO and mitigate the competing adsorption of *H in the reaction process.

The family of noble metal dioxides (such as $RuO_2$ and $IrO_2$) is a well-known catalyst and has been employed in a widely heterogeneous catalytic reaction including the oxidation of CO and $NO_x$, the gradation of alcohols, and dehydrogenates $NH_3$, anodic evolution of chlorine in Deacon process and so on[129-133]. More interestingly, the $IrO_2$ and $RuO_2$ binary oxides recently have been confirmed to have Dirac nodal lines and exhibit a large spin Hall effect or a crystal Hall effect[134-136]. The surface adsorption activities of $H_2$, $O_2$, NO, and CO in the active (110) surface of $RuO_2$ have been studied by in operando APRES. It is found that $RuO_2$ (110) can oxidize CO and $H_2$, but can't afford to oxidize NO, implying the active role of the flat-band surface states in catalytic charge transfer processes at the oxygen bridge sites[137]. This approach further provides helps to search for more effective topological catalysts. However, this method still has limitations. Whether topology and/or correlation physics such as magnetism impact surface activities of TQMs, is still hard to be detected by ARPES experimentally so far. Distinguished from the DSMs (like $Na_3Bi$ and $Cd_3As_2$), the cubic antiperovskite material $Cu_3PdN$ has been recently predicted to be a 3D nodal line semimetal[138] and a possible topological superconductor[139]. $Cu_3PdN$ with rich physical properties also has a superior electrocatalytic activity for $CO_2$ reduction to formic acid to the non-topological compounds $Cu_3N$ and $Cu_3Pd$ (REF.[140]). In addition, $Cu_3PdN$ is a potential candidate for oxygen reduction reaction (ORR) under alkaline conditions[141]. The effect of TSSs in $Cu_3PdN$ on these heterogeneous catalytic reactions has not been considered and uncovered for this moment. Recently, 2D topological metal monolayer $\beta$-$PdBi_2$ is also reported to show outstanding catalytic performance for $CO_2$ electro-reduction to formic acid, where the additional electronic states at the Fermi level caused by the inherent SOC effect remarkably improve the adsorption of OCHO* intermediate on $\beta$-PdBi$_2$ monolayer, resulting in the low potential of -0.26 V vs. reversible hydrogen electrode (RHE) for CO$_2$ electro-reduction to formic acid (REF.[124]). APRES has been recently used to characterize the Dirac nodal-line fermions in the Cu$_2$Si monolayer (REFS[142,143]). Accordingly, 2D topological Cu$_2$Si nanoribbons with a chair form edge (marked as A > CuSi) and a serrated edge winding-up of Cu or Si (marked as Z > Cu and Z > Si) can reduce carbon dioxide to *COOH with small barriers due to the edge-state-enhanced conductivity in NLSMs. Specifically, S > Si is most active for further hydrogenation, exhibiting the capability of transporting eight electrons to yield methane with a small free energy variation of 0.24 eV (Fig. 4a). For CO$_2$ conversion reactions, the side reaction of HER should be taken into consideration as it is generally more thermodynamically favorable. It is interesting to see that DNLs Cu$_2$Si exhibited high selectivity towards the target reaction of CO$_2$ reduction. As displayed in Fig. 4b, the free energy change of HER is 0.36 eV at the Z > Si site, which is significantly larger than the target reaction with the value of 0.24 eV(REF.[144]). This suggests that the CO$_2$ reaction is more thermodynamically favorable and the side reaction can be inhibited efficiently.

**Organic molecules conversions**

The superconductor and type-II Dirac semimetal, PdTe$_2$ has been used for ethanol electrooxidation (EOR) and the mass activity of PdTe$_2$ nanosheet electrocatalyst is increased by about quintupling as compared with that of commercial Pd black (**Fig. 4c**). Moreover, the forward current (J$_f$) and backward current (J$_b$) of PdTe$_2$ electrocatalyst in the proportion of 5.28:1, which is much larger than previously observed (J$_f$/J$_b$ ≈ 2), suggesting outstanding oxidation completeness of ethanol. Besides, PdTe$_2$ topological electrocatalyst exhibits a low Tafel slope (41.2 mV dec$^{-1}$) and comparable durability (see **Fig. 4d**)[10].

Selective hydrogenation is a significant reaction in the pharmaceutical and petrochemical fields. It is of particular importance in the polyethylene production processes, where the removal of alkyne impurities from alkene feedstock is vital to inhibit the poisoning of the heterogeneous catalyst during alkene polymerization. Recently, some TQMs (such as PtGa chiral crystal, $Co_2Mn_xFe_{1-x}Ge$ Heusler alloys) also have been considered as new catalysts for the selective hydrogenation of alkynes based on the conventional $d$ band theory with the Brønsted-Evans-Polanyi (BEP) relation[125,145]. Yet, there is a lack of discussions on the relationship between TSSs and catalytic performance.

As shown in the aforementioned section on water splitting, the TQMs with robust TSSs and high mobility are in favour of the HER and OER processes. In addition, recently, it has been found that the TQMs with robust conductivity protected by topology also can keep in the storage of hydrogen. The excellent hydrogen storage performance in $Li_2CrN_2$ (REF.[146]) (**Fig. 4e**) sheet with a gravimetric capacity of 4.77% is observed (**Fig. 4f**), which is superior to the Li decorated $MoS_2$ (4.4%) and Li-decorated phosphorene (4.4%). Besides, the hydrogen adsorption energy within the area of 0.16 - 0.33 eV is in the requisite energy area for the balance of the adsorption stability and fast kinetics, meanwhile, the releasing temperature within the area of 160 - 270 K is preferable for practical usages[147].

TQMs have been rationally designed as robust catalysts for the selective hydrogenation and oxidation of biomass-derived alcohols (like furfuryl alcohol and benzyl alcohol), acid (such as levulinic acid), and aldehydes (for example, furfural, cinnamaldehyde, and benzaldehyde). Fine Pd, Pt, Cu nanoparticles (less than 10 nm) mainly exhibited (111) crystal plane have exhibited good performance in the hydrogenation of biomass-derived levulinic acid and furfural, whereas biofuel component gamma-valerlactone (over 98% yield) and 2-methyl furan (over 55%) were selectively produced. This superior performance is closely related to the tempting electronic properties of TQMs as well as the unsaturated $d$-band in enhancing the adsorption of reactants and hydrogen[148-150].

# Outlook

Topological quantum catalysts use the topologically protected surface bands as active sites, which is different from the edge-, vacancy-, dopant-, strain-, or heterostructure-created sites. This will allow us to understand and identify the origin of catalytic activities. The topological properties of TQMs also meet the criteria of high-performance battery materials, supercapacitors, and thermoelectrics, and thus they have been received increasing research efforts. Thanks to the rapid progress in topological theories such as topological quantum chemistry, the high-throughput first-principles calculations are powerful in searching and identifying topological materials, with several databases have been established[4,151,152]. Researchers can identify the topological properties and band structures fast and precisely from the databases now, although improvements are still needed.

Even though the research on TQMs for energy conversion and storage is developing rapidly, this field is still immature, especially in terms of their relationships between topological states and energy conversion as well as storage processes derived from topological states. Unlike the exploration of topological properties such as resistivities and various Hall effects, the chemical reactions are generally happening in critical conditions such as acidic, alkaline, and even high vacuum. This makes the in-situ observation of topological effects an extremely difficult task now. To find direct evidence, high-quality bulk single crystals and in-situ monitoring of the band topology evolution using spectroscopy are urgently needed. A good example is the study on the extensively investigated $MoTe_2$. With non-topological 2H-$MoTe_2$ and topological 1T'-$MoTe_2$, it is possible to isolate the contrition from the topological phase by tracking the change of surface compositions, conductivities, carrier mobilities, and Hall resistivities before and after electro/photocatalytic reactions[153]. In addition, in-situ tools (such as EXAFS, TEM, nano-ARPES) combined with theoretical calculations are highly desirable. It has been preliminary confirmed that ARPES is powerful to track the response of bulk DSMs and TSSs to redox reactions in operando, although the high vacuum is still necessary[137]. Luckily, routine angle-resolved experiments at near-ambient conditions might become the most advanced technologies in the long term. Invested with extra time and spin resolution, an in-situ inspection of catalytic spin transfer and further clarification of the effect of surface magnetism on heterogeneous catalytic reactions can be achieved[154-156].

Another challenge is to find a TQM catalyst that is suitable for industrial-scale reactions. For hydrogen production and fuel cell reactions, the catalysts are expected to work on an industrial scale for a long time. Unfortunately, kinetic and stability assessments of most catalysis and conversion reactions focus on the low current densities (10 mA cm$^{-2}$) and potential ranges. Increasing the applied potential may destroy the surface of catalysts, or require high energy consumption that is far beyond what we can bear. Considering that both the charge and spin of electrons could be transported simultaneously for all the redox reactions, it is possible to boost the catalysis efficiency through the radical pair effects by manipulating spin polarization. Previous research has partially confirmed that the spin-polarized topological states can indeed interfere with the catalytic process directly[8,157], and most importantly, the spin polarization can be tuned by external fields such as current or magnetic fields[158,159]. In the presence of a commercially permeated magnet, an improvement of 150 % in the OER current density can be obtained because of the polarized electrons in magnetic catalysts[160]. Similarly, NbP family of WSMs exhibits a positive response at a magnetic field of 0.35 T. As hydrogen evolution catalysts, the efficiency is increased by 11 % for NbAs and 95 % for NbP[161]. More interestingly, the negative effects of surface oxidation and re-construction can be overcome as the spin properties can be reserved through the spin pinning effect[150]. Although positive effects are reported in the presence of the magnetic field, we find that the intrinsic catalytic activities can be suppressed significantly, for example, the turnover frequency of topological semimetal $Co_2MnGa$ is decreased by 33 % as HER catalysts[154]. Thus, how the magnetic fields and spin polarization 'talk' with the chemical reactions requires more insightful exploration.

As for energy storage devices including the LIBs, NIBs, KIBs, and supercapacitors, the electrode materials are one of the most vital factors to realize the high specific capacity, high energy density, and excellent cycling stability. TQMs with high electrical conductivity and desirable carrier mobility are the promising candidate electrode materials for these energy storage devices. Yet, the synthesis of promising topological electrode materials is another major challenge. Most of the topological batteries are still in theoretical perspective, synthetic condition even needs more efforts[99]. Moreover, the fundamental knowledge such as the suitable voltage windows and the discharge voltage platform of different TQM electrode materials in the different energy storage devices is unclear yet. These issues need the considerable effort of researchers to be devoted. However, we are confident that in near future, except for these applications in this perspective, more and more new energy-relative applications such as biomass conversion, nitrogen fixation will be attempted and achieved, in which chemists and physicists need to make joint efforts.

**Acknowledgments**

This work is supported by the National Natural Science Foundation of China (11922415, 22078374, 21776324), Guangdong Basic and Applied Basic Research Foundation (2022A1515011168, 2019A1515011718, 2019B1515120058, 2020A1515011149), Key Research & Development Program of Guangdong Province, China (2019B110209003), the Pearl River Scholarship Program of Guangdong Province Universities and Colleges (20191001) and Hundred Talent Plan from Sun Yat-sen University, and the Foundation of President of Ningbo Institute of Materials Technology and Engineering (NIMTE) of the Chinese Academy of Sciences (CAS).

**Author contributions**

H. Luo is the leading author in organizing this perspective. All authors occupy the writing and polish the manuscript.

**Competing interests**

The authors declare no competing interests.

**Peer review information**

Nature Reviews Physics thanks Ankita Anirban and the other, anonymous, reviewer for their contribution to the peer review of this review.

**Publisher's note**

Springer Nature remains neutral with regard to jurisdictional claims in published maps and institutional affiliations.

# Table 1

**Table 1** | HER and OER performance for various topological quantum materials (TQMs), where the TIs, TSMs, and TSCs represent the topological insulators, topological semimetal, and topological superconductor, respectively.

| | | | Water splitting | | | | |
|---|---|---|---|---|---|---|---|
| Applications | TQMs | Compounds | Conductivity ($\sigma$)/mobility ($\mu$) value at room temperature | Morphology | $\eta_{10}$ mA cm$^{-2}$ (mV) | Tafel slope (mV dec$^{-1}$) | Stability |
| HER | TIs | Pd/SnTe (REF.[41]) | $\sigma_{SnTe} = 8.06 \times 10^3$ S·cm$^{-1}$ $\mu_{SnTe} = 46$ cm$^2$·V$^{-1}$ s$^{-1}$ | Metal/TI Heterostructure | 86 | 29.14 | 1000 CV cycles |
| | | Bi$_2$Te$_3$ (REF.[38]) | $\sigma = 1.22 \times 10^3$ S·cm$^{-1}$ $\mu = 28$ cm$^2$·V$^{-1}$ s$^{-1}$ | Thin film | 219 | 47.87 | 48 h at 10 mA cm$^{-1}$ |
| | | Bi$_2$Te$_3$ (REF. [44]) | $\mu >10^3$ cm$^2$ V$^{-1}$ s$^{-1}$ | nanoparticles | - | - | - |
| | TSMs | NbP (REF.[62]) | | Single crystal | 3520 μmol g$^-$ | - | - |

|  |  |  |  |  |  |
|---|---|---|---|---|---|
| | $\sigma = 6.30 \times 10^5$ S·cm$^{-1}$ | | $^1$ for 6.3 h | | |
| NbP (REF.$^{61}$) | $\mu = 2.75 \times 10^2$ cm$^2$·V$^{-1}$·s$^{-1}$ | Single crystal powder | 7442 µmol g$^{-1}$ for 6 h | - | - |
| NbP (REF.$^{160}$) | | Single crystal powder | 333 µmol g$^{-1}$ for 3 h | - | - |
| TiSi (REF.$^{11}$) | $\sigma = 3.23 \times 10^4$ S·cm$^{-1}$ | Polycrystal | 34 | 27.1 | 1000 CV cycles |
| MnSi (REF.$^{11}$) | $\sigma = 8.17 \times 10^3$ S·cm$^{-1}$ | Polycrystal | 39 | 27.3 | 1000 CV cycles |
| FeSi (REF.$^{11}$) | $\sigma = 2.6 \times 10^4$ S·cm$^{-1}$ | Polycrystal | 42 | 29.7 | 1000 CV cycles |
| RuSi (REF.$^{11}$) | $\sigma = 1.11 \times 10^2$ S·cm$^{-1}$ | Polycrystal | 42 | 23.6 | 1000 CV cycles |
| NiSi (REF.$^{11}$) | $\sigma = 9.65 \times 10^4$ S·cm$^{-1}$ | Polycrystal | 43 | 30.1 | 1000 CV cycles |
| CoSi (REF.$^{11}$) | $\sigma = 7.47 \times 10^3$ S·cm$^{-1}$ $\mu = 119$ cm$^2$·V$^{-1}$·s$^{-1}$ | Polycrystal | 54 | 31.5 | 1000 CV cycles |
| RhSi (REF.$^{11}$) | $\sigma = 3.61 \times 10^3$ S·cm$^{-1}$ | Polycrystal | 114 | 50.8 | 1000 CV cycles |
| 1T'-MoTe$_2$ (REF.$^{64}$) | $\sigma = 1.91 \times 10^3$ S·cm$^{-1}$ $\mu = 16.2$ cm$^2$·V$^{-1}$·s$^{-1}$ | Bulk form | 356 | 127 | 1 h |
| PtTe$_2$ (REF.$^{48}$) | $\sigma = 3.3 \times 10^6$ S·cm$^{-1}$ $\mu = 4.41 \times 10^2$ cm$^2$·V$^{-1}$·s$^{-1}$ | Layered bulk form | 540 | 110 | - |
| MoP (REF.$^{163}$) | $\sigma = 4.30 \times 10^5$ S·cm$^{-1}$ $\mu = 2.32 \times 10^3$ cm$^2$·V$^{-1}$·s$^{-1}$ | Nanocrystals | 154 | 102 | - |
| WTe$_2$ (REF.$^{164}$) | $\sigma = 7.41 \times 10^2$ S·cm$^{-1}$ $\mu = 84.21$ cm$^2$·V$^{-1}$·s$^{-1}$ | Nanoribbons | 430 | 57 | 20 h at 10 mA cm$^{-2}$ |
| PdGa (REF.$^{16}$) | $\sigma = 6.67 \times 10^4$ S·cm$^{-1}$ | Single crystal | 207 | 120 | - |
| RhSi (REF.$^{16}$) | $\sigma = 3.61 \times 10^3$ S·cm$^{-1}$ | Single crystal | 280 | 145 | - |
| PtGa (REF.$^{16}$) | $\sigma = 1.08 \times 10^4$ S·cm$^{-1}$ | Single crystal | 13.3 | 16 | - |
| PtAl (REF.$^{16}$) | - | Polycrystalline ingot | 14 | 20 | 100 h at different |

| | |  |  |  |  |  |  |
|---|---|---|---|---|---|---|---|
| | | | | | | | overpotential |
| | | PtSn$_4$ (REF.[13]) | $\sigma = 1.27 \times 10^2$ S·cm$^{-1}$<br>$\mu = 200$ cm$^2$·V$^{-1}$·s$^{-1}$ | Single crystal | 37 | 39 | 400 h at 46 mV |
| | | PtTe$_2$ (REF.[165]) | $\sigma = 3.3 \times 10^6$ S·cm$^{-1}$<br>$\mu = 4.41 \times 10^2$ cm$^2$·V$^{-1}$·s$^{-1}$ | Thin film | 330 | 85 | - |
| | | 1T'-MoTe$_2$ (REF.[64]) | $\sigma = 1 \times 10^3$ S·cm$^{-1}$<br>$\mu = \times 10^2$ cm$^2$·V$^{-1}$·s$^{-1}$ | Single crystal | 73 | 46.3 | 10 h |
| | | VAl$_3$ (REF.[63]) | $\sigma = 8.64 \times 10^3$ S·cm$^{-1}$<br>$\mu = 1.31 \times 10^2$ cm$^2$·V$^{-1}$·s$^{-1}$ | Microscale single crystals | 319 | 68 | - |
| | | V$_{0.75}$Ni$_{0.25}$Al$_3$ (REF.[63]) | $\mu = 4.67$ cm$^2$·V$^{-1}$·s$^{-1}$ | Microscale single crystals | 175 | 73 | 30 h at 200 mV |
| | Possible TSCs | PdTe$_2$ (REF.[10]) | $\sigma = 4.30 \times 10^6$ S·cm$^{-1}$<br>$\mu = 1.19 \times 10^3$ cm$^2$·V$^{-1}$·s$^{-1}$ | Layered bulk form | 740 | 92 | - |
| | | α-BiPd (REF.[9]) | $\sigma = 2.10 \times 10^4$ S·cm$^{-1}$ | Nanosheets | 144 | 104 | 13 h at 500 mV |
| OER | TSMs | Co$_3$Sn$_2$S$_2$ (REF.[12]) | $\sigma = 2.68 \times 10^3$ S·cm$^{-1}$<br>$\mu = 2.6 \times 10^3$ cm$^2$·V$^{-1}$·s$^{-1}$ | Single crystal | 300 | 74 | 12 h at 1.497 mV |
| | | Co$_2$MnAl (REF.[83]) | $\sigma = 4.12 \times 10^3$ S·cm$^{-1}$ | Crystal | - | 68 | - |
| | | Co$_2$MnGa (REF.[83]) | $\sigma = 5.77 \times 10^3$ S·cm$^{-1}$ | Crystal | - | 67 | 12 h at 1.63 V |
| | | NiPS$_3$ (REF.[166]) | $\sigma = 1 \times 10^{-7}$ S·cm$^{-1}$<br>$\mu = 3.5$ cm$^2$·V$^{-1}$·s$^{-1}$ | Nano sheets | 301 | 43 | 20 h at 10 mA cm$^{-2}$ |

# Table 2

**Table 2** | Batteries and supercapacitors performance for various topological quantum materials (TQMs), where the TIs, TSMs, and TSCs represent the topological insulators, topological semimetal, and topological superconductor, respectively.

| \multicolumn{7}{c}{Battery/ Supercapacitors} |||||||
|---|---|---|---|---|---|---|
| Applications | TQMs | Compounds | Conductivity ($\sigma$)/mobility ($\mu$) value at room temperature | Morphology | Specific capacity (mAh g$^{-1}$) | Maximum energy density (Wh kg$^{-1}$) | Stability |
| LIBs | TSMs | bco-C16 (REF.[87]) | - | Bulk form | 558 | - | - |
| | | HZMG-42 (REF.[94]) | $\sigma = 1 \times 10^{-6}$ S·cm$^{-1}$ | Bulk form | 638 | - | - |
| | | m-C16 (REF.[88]) | - | Bulk form | 558 | - | - |
| NIBs | TSMs | tC24 (REF.[90]) | - | Bulk form | 233 | - | - |
| | | $T_d$-WTe$_2$ (REF.[105]) | $\sigma = 7.41 \times 10^2$ S·cm$^{-1}$ $\mu = 84.21$ cm$^2$·V$^{-1}$ s$^{-1}$ | Layered bulk form | 245 (0.1 A g$^{-1}$) | - | - |
| | | NaAlSi (REF.[91]) | $\sigma = 6.02 \times 10^2$ S·cm$^{-1}$ | - | 476 | - | - |
| | | NaAlGe (REF.[91]) | - | - | 238 | - | - |
| KIBs | TSMs | $T_d$-WTe$_2$ (REF.[105]) | $\sigma = 7.41 \times 10^2$ S·cm$^{-1}$ $\mu = 84.21$ cm$^2$·V$^{-1}$ s$^{-1}$ | Layered bulk form | 238 (0.1 A g$^{-1}$) | - | - |
| ZIBs | TSMs | Co$_3$Sn$_2$S$_2$ (REF.[114]) | $\sigma = 2.68 \times 10^3$ S·cm$^{-1}$ $\mu = 2.6 \times 10^3$ cm$^2$·V$^{-1}$ s$^{-1}$ | Crystal pellets | 160 (1 A g$^{-1}$) | 305 (0.2 A g$^{-1}$) | 1460 cycles at 1 A g$^{-1}$ |
| \multicolumn{7}{c}{Supercapacitors} |||||||
| Applications | TQMs | Compounds | Conductivity ($\sigma$)/mobility ($\mu$) value at room temperature | Morphology | Specific mass capacitance (F g$^{-1}$) | Maximum energy density (Wh kg$^{-1}$) | Stability |
| Supercapacitors | TSMs | $T_d$-WTe$_2$ (REF.[115]) | $\sigma = 7.41 \times 10^2$ S·cm$^{-1}$ $\mu = 84.21$ cm$^2$·V$^{-1}$ s$^{-1}$ | Thin film | 221 (1 A g$^{-1}$) | 31 | 5500 cycles at 3.4 A cm$^{-3}$ |

# Figures

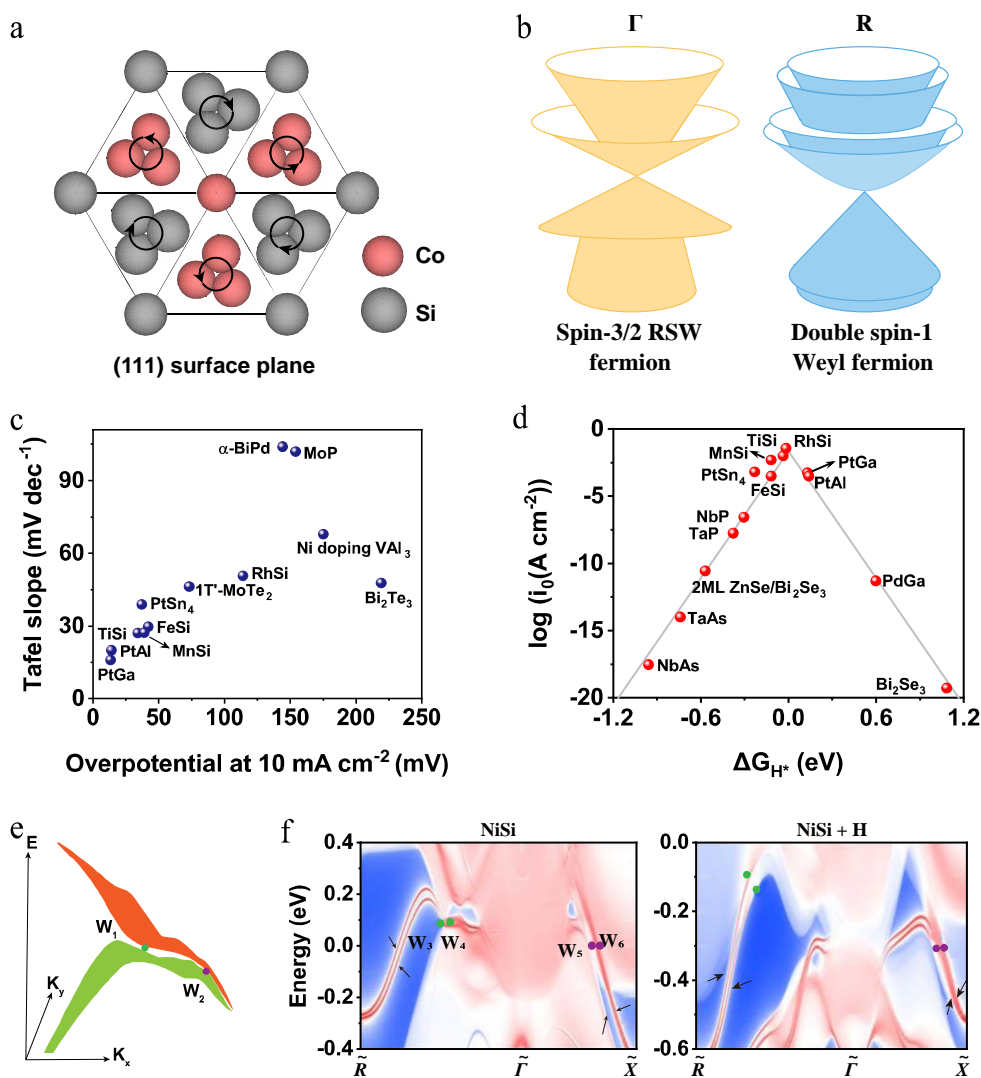

Fig. 1 | **Topological quantum materials for HER. a** | Schematic top view of (111) surface of topological chiral semimetal CoSi. The anticlockwise and clockwise circles indicate the chirality of the Co and Si atoms, respectively[167]. **b** | The fourfold degenerated spin-3/2 RSW fermion node at the Γ point and the sixfold degenerated double spin-1 Weyl nodes at the R point for CoSi. coupling along the high-symmetry directions. **c** | Scatter diagram of Tafel slopes and overpotentials at 10 mA cm$^{-2}$ for metal monosilicides and other typical topological materials. Tafel slope indicates the charger transfer ability and the lower Tafel slope means the faster charge transfer ability. While the overpotential implies the extra potential for overcoming the energy

barrier to initiate the reaction. These data are collected from REFS[9,11,13,16,38,63,64,162]. **d** | Volcano plot of various catalysts at a certain surface. These data are collected from REFS[11,13,16,40,77]. **e** | The hybrid Weyl nodes by 3D plotting of band dispersions for NiSi. **f** | The comparison of (001) surface states of NiSi before and after the hydrogen adsorption. Panel **b** reprinted with permission from REF.[75], CC BY-NC (http://creativecommons.org/licenses/by-nc/4.0/legalcode). Panel **e, f** reprinted with permission from REF.[81], Elsevier.

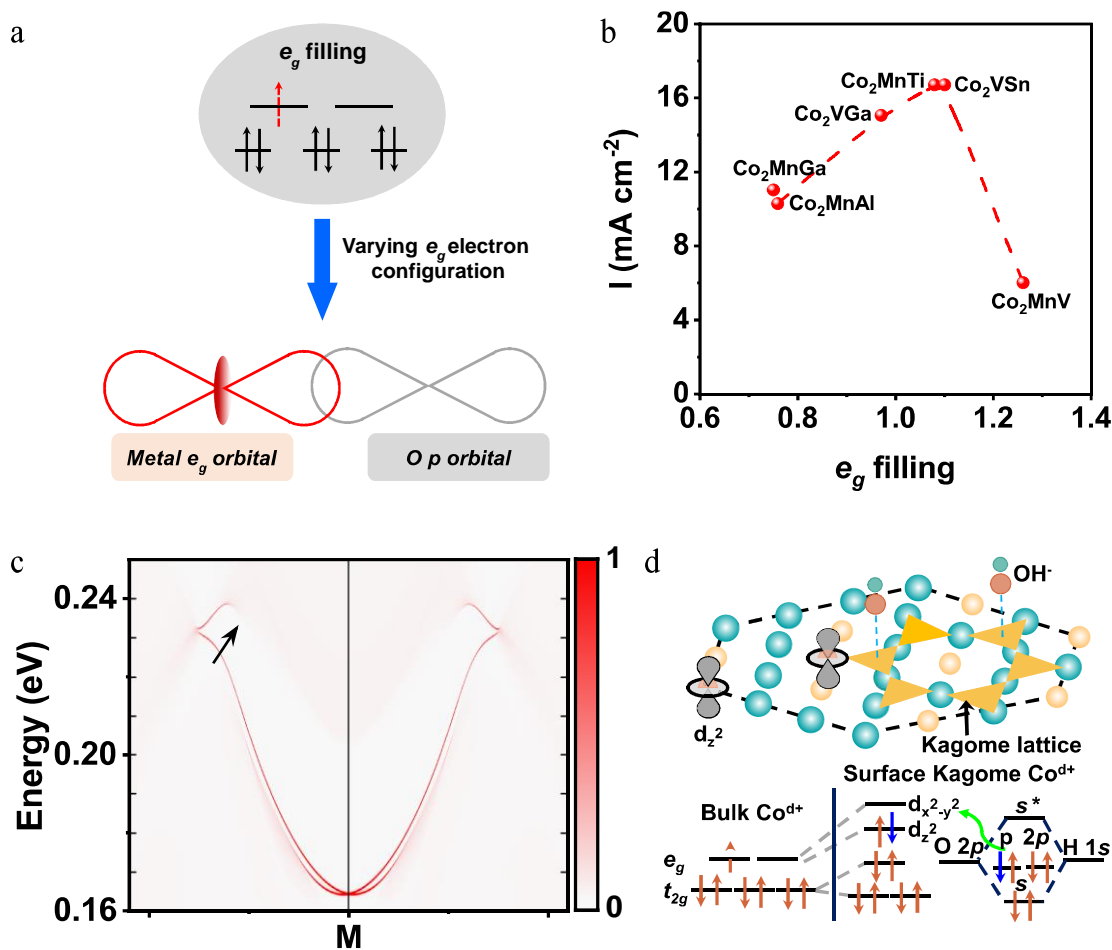

Fig. 2 | **Topological materials for OER. a** | According to the Shao-Horn (SH) principle, the binding energy of oxygen intermediates to the catalyst surface can be affected by the $e_g$ orbital filling, and thus the OER activity can be altered by varying the eg electroon configuration. **b** | The volcano plot of the OER performance, which is derived from the current density at 1.7 V vs. RHE against the occupancy of the $e_g$ electron of Co in Heusler compounds ($Co_2MnZ$ and $Co_2VZ$). **c** | The nontrivial surface states on (001) plane of $Co_3Sn_2S_2$ crystal with Sn termination, which is not completely occupied and located just above the Fermi level. **d** | Schematic diagram of favourable OH uptake with the Co $3d$ orbitals. Panel **b** reprinted with permission from REF.[83], CC BY 4.0 (https://creativecommons.org/licenses/by/4.0). Panel **c,d** reprinted with permission from REF.[12], CC BY 4.0 (https://creativecommons.org/licenses/by/4.0).

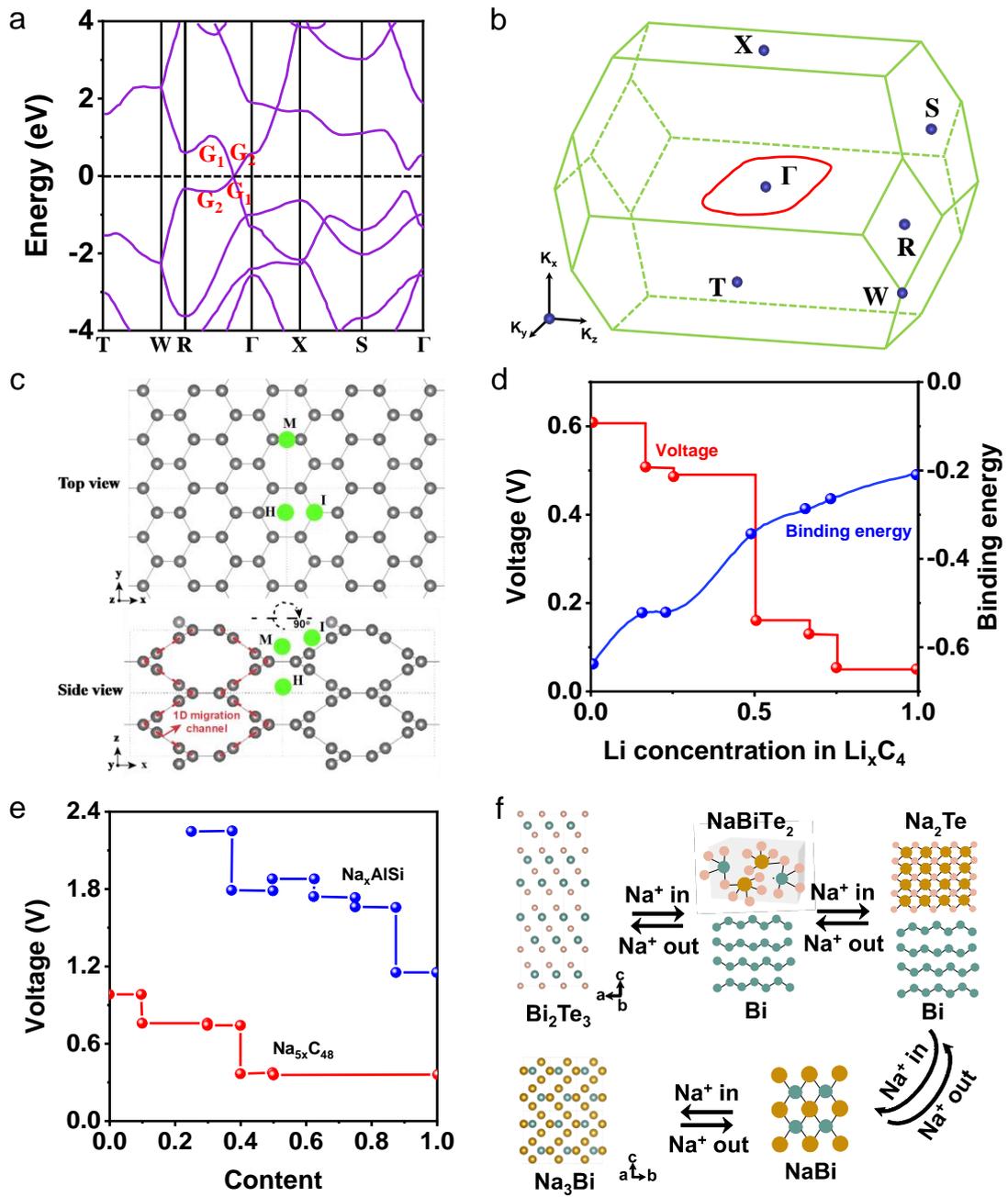

Fig. 3 | **Topological materials for battery and supercapacitors. a** | The bulk band structure along several high-symmetry directions for bco-$C_{16}$. The G1 and G2 represent the irreducible representation of the two crossing bands, respectively. **b** | The Brillouin zone of bco-$C_{16}$ with several high-symmetry momenta. The red circle indicates the nodal ring. **c** | Structure of bco-C16 and the schematics of the possible Li-ion absorption sites in top and side view. **d** | The voltage profile and binding energy profile are calculated along the minimum energy path. **e** | The voltage profile different Na contents for four stable intermediate phases in $Na_{5x}C_{48}$ and the average voltage profile for

Na$_x$AlSi ($x$ = 0.25, 0.375, 0.5, 0.625, 0.75, 0.825, 1.0) with disparate Na contents. There is a large voltage plateau (0.98 V) for $0 < x < 0.1$, which may be due to the forceful binding of Na with tC$_{24}$; the voltage is maintained around 0.75 V when $0.1 < x < 0.4$; there is an obvious decline from 0.74 to 0.37 V at Na$_{0.4}$tC$_{24}$ when $0.4 < x < 1.0$, as a result of the raising repulsive interaction between the Na ions. These data are collected from REFS[90,91]. **f** | Schematic of the phase-change mechanism of Bi$_2$Te$_3$ during sodiation/desodiation process. These data are collected from REF.[104]. Panel **a, b** reprinted with permission from REF.[89], APS. Panel **c, d** reprinted with permission from REF.[87], National Academy of Sciences.

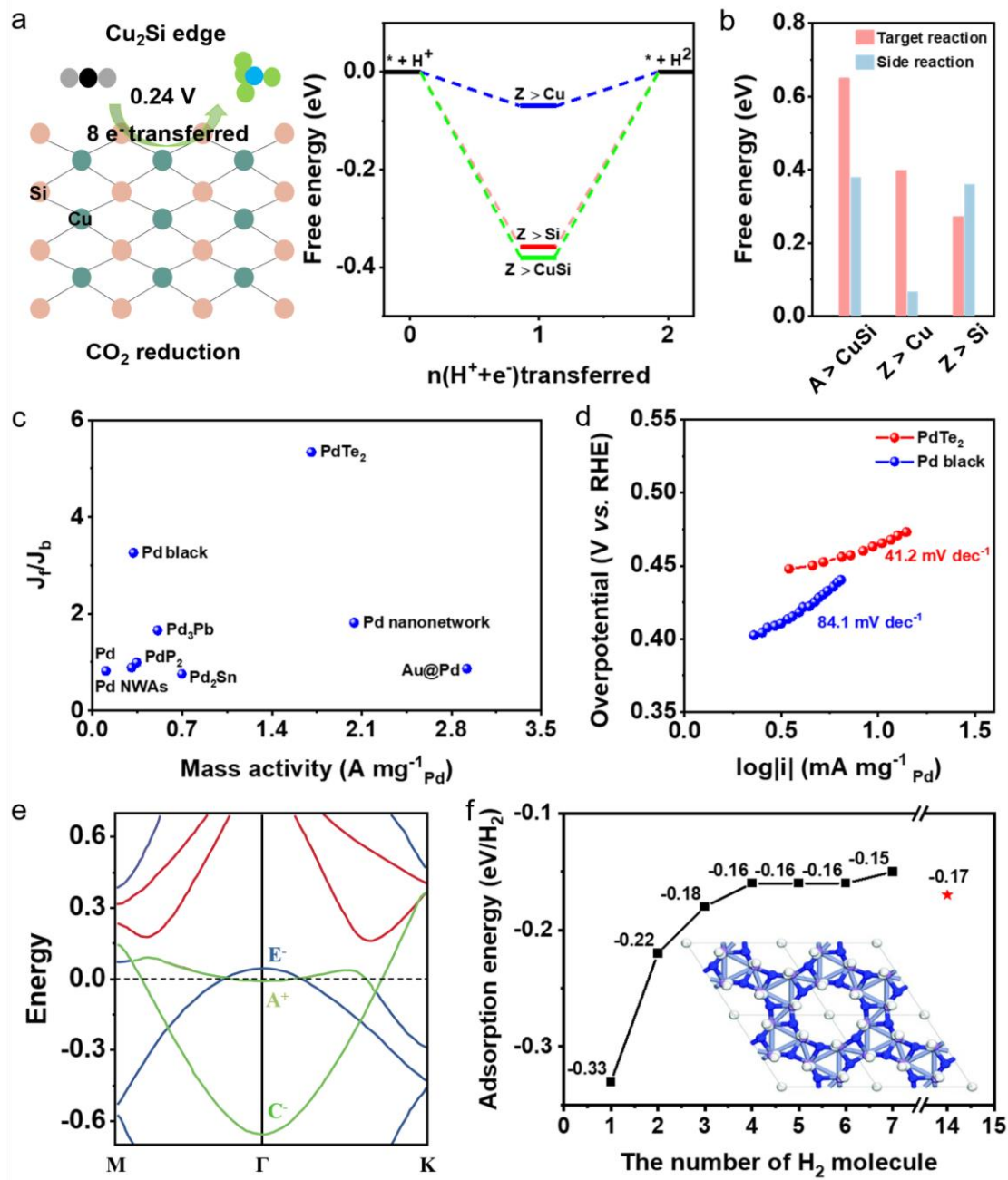

Fig. 4 | **Topological materials for other applications. a** | Illustration of $CO_2$ electro-reduction on topological nodal line semimetal $Cu_2Si$ and side reaction profiles of nanoribbons. **b** | The comparisons of free energy change between the target reaction and the competition reaction. **c** | The comparisons of the ratio of forwarding current to backward current and mass activities between $PdTe_2$ nanosheets and other typical catalysts. **d** | Tafel plots of the $PdTe_2$ nanosheets and Pd black measures in a 1 M KOH + 1 M EtOH solution and the overpotential and current density are obtained from linear sweep voltammetry curves. **e** | The orbital-projected band structure of $Li_2CrN_2$ at a

strain of −10%. **f** | Changes of adsorption energies $E$(PBE+TS) with various hydrogen molecules. The optimized geometry with maximum hydrogen storage capacity is calculated according to the G = 14 × M(H$_2$)/(M(Li$_2$CrN$_2$) + 14 × M(H$_2$)), where the sign of M presents the mass. Panels **a-b** reprinted with permission from REF.[144], American Chemical Society. Panels **c-d** reprinted with permission from REF.[10], Wiley. Panel **e** reprinted with permission from REF.[146], Springer Nature. Panel **f** reprinted with permission from REF.[147], Royal Society of Chemistry.

**Box 1 General diagram for catalysis and energy conversion**

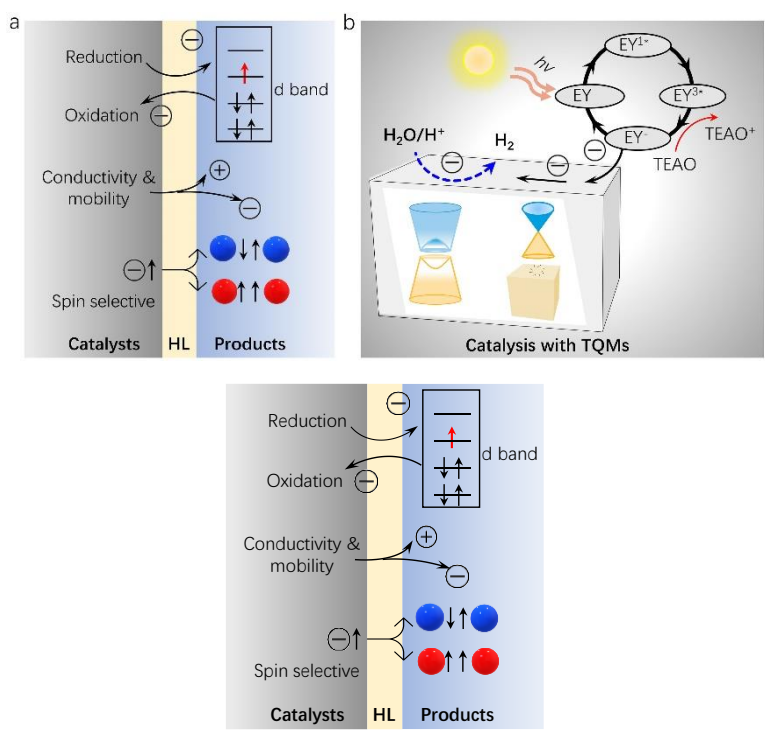

Understanding chemical bonds and electron transfer are essential to describing catalysis theory from physical and chemical aspects[168,169]. One of the most successful theories to describe catalysis is called the d-band model because $d$ orbital electrons of the catalysts play a vital role in the process[170]. Various linear relationships based on the $d$-band model have been proposed to understand the catalysis mechanisms and then predict new high-performance catalysts, such as adsorption energies of key reaction intermediates, $d$ orbital filling numbers, and electronic spin moment[171-173]. A volcano plot can be then created from the linear relationships, which states that there is always a preferred value when the catalysis exhibits the maximum activities. For example, the Gibbs free energy of hydrogen should be neither too strong nor too weak for a good HER catalyst.

Electron transfer is an essential step for redox catalysis reactions (see panel). The essence of catalysis reactions is the electron transfer process between the catalysts and the reactants, either from the catalysts to reactants (oxidation), or vice versa (reduction). Thus, as two of the most important properties of electrons, both charge and spin should be taken into consideration. Most catalysts require high electrical conductivities to decrease the charge transfer resistances across the bulk and interface[174,175]. High carrier mobility is another important indicator for promising catalysts. The desirable carrier mobility guarantees a fast utilization of the excited electrons for redox reactions, rather than being trapped by defects or recombination with holes[176,177].

In addition to conductivity and mobility, electron spin properties are often overlooked in understanding heterogeneous catalysis reactions. In fact, nearly all reaction steps, from the bond formation adsorbate, to electron transfer, can be manipulated by spin. This is because various radicals are formed during the reactions, and their reduction or oxidation is strongly dependent on the spin states of the accepted or donated electron from the catalysts. In other words, the spin states of the radical-electron pair are closely related to the electron transfer kinetics and catalytic conversion efficiencies. Such a radical-pair mechanism opens the field of spin chemistry and highlights the importance of spin electrons in high-performance designing[178,179].